\documentstyle[preprint,prd,aps]{revtex}

\def\gsim{\begin{array}{c} > \\ \sim \end{array}}
\def\lsim{\begin{array}{c} < \\ \sim \end{array}}

\begin{document}
\draft
\preprint{\vbox{\hbox{gr-qc/9911054} }}
\title{\large \bf On Wigner's clock and the detectability of spacetime
foam with gravitational-wave interferometers}
\author{\bf Y. Jack Ng$^{(a),(b)}$\thanks{Corresponding author. E-mail:
yjng@physics.unc.edu}
and H. van Dam$^{(b)}$}
\address{(a) Center for Theoretical Physics, Laboratory for Nuclear
Science and Department of Physics,\\
Massachusetts Institute of
Technology,
Cambridge, MA 02139\\}
\address{(b) Institute of Field Physics, Department of Physics
and Astronomy,\\
University of North Carolina, Chapel Hill, NC 27599-3255\\}
\maketitle
\bigskip

\begin{abstract}

A recent paper (gr-qc/9909017) criticizes 
our work on the structure of spacetime foam.  Its authors argue
that the quantum uncertainty limit for the position of the quantum
clock in a gedanken timing experiment, obtained by Wigner and used by us,
is based on unrealistic
assumptions.  Here we point out some flaws in their argument.  We also
discuss their other comments and some other issues related to
our work, including a simple connection to the holographic principle.  We
see no
reason to change our cautious optimism on the detectability of spacetime
foam with future refinements of modern gravitational-wave interferometers
like LIGO/VIRGO and LISA. 


\end{abstract}

\newpage
\section{Introduction}

In their recent paper\cite{anos} "On the detectability of quantum
spacetime foam with
gravitational-wave interferometers," Adler, Nemenman,
Overduin,
and Santiago claim that the way we use Wigner's quantum clock
\cite{wigner} in
a gedanken timing experiment is not justified, thus casting doubt on the
detectability of spacetime foam with gravitational-wave interferometers.
In particular, they claim that the quantum uncertainty limit for the
position of the quantum clock is actually much smaller than that obtained
by Wigner and used by us.  Since we \cite{nvd1,nvd2} were the first to
propose using Wigner's clock to explore the
quantum structure of spacetime and to conclude that classical spacetime
breaks down into "quantum foam" in a manner quite different from the
canonical picture \cite{MTW}, we feel a special obligation to respond to
the criticism and to clarify the physics behind our proposal.  We will
show that the arguments by Adler et al. are invalid.

But first, we should make it clear that we merely want to find out what
the low-energy limit of quantum gravity can tell us about the structure of
spacetime.  For that purpose, it suffices to employ the general principles
of quantum mechanics and general relativity.  We have little to say about,
and for this work, have no use for,
the correct theory of quantum gravity (be it string theory, Ashtekar
variables/loop-gravity formalism, or something else).  We have in mind the 
low-energy
limit of quantum gravity as manifested in the low-frequency spectrum of
the displacement noise levels registered in the gravitational-wave
interferometers.

In the next section, we recapitulate our previous work 
\cite{nvd1,nvd2,nvd3} 
on spacetime measurements and spacetime foam. In Section III,
we respond to each of the four objections against our work raised by Adler
et al.  In Section IV, we answer some further questions which we think the
readers may ask.  We offer our conclusions in Section V.  We point out that
our spacetime foam model\cite{nvd1,nvd2,nvd3} is consistent with the
holographic
principle\cite{tHooft}.

\section{Space-time measurements and the foaminess of spacetime }

Suppose we want to measure the distance between two separated points A and
B.  To do this, we put a clock (which also serves as a light-emitter and
receiver) at A and a mirror at B.  A light signal is sent from A to B
where it is reflected to return to A.  
If the clock reads zero when the light signal is emitted and reads $t$
when the signal returns to A, then
the distance between A and B is given by $l = ct/2$, where $c$
stands
 for the speed of light.  The next question is: What is the
uncertainty (or error) in the distance measurement?  Since the clock at A
and the mirror at B are the agents in measuring the distance, the
uncertainty of distance $l$ is given by the uncertainties in their
positions.  We will concentrate on the clock, expecting that the mirror
contributes a comparable amount to the uncertainty in the measurement of
$l$.
Let us first recall that the clock is not stationary; its spread in speed at

time zero is given by
the Heisenberg uncertainty principle as
\begin{equation}
\delta v =  \frac{\delta p}{m} \gsim \frac{\hbar}{2m\delta l},
\label{ineq1}
\end{equation}
where $m$ is the mass of the clock.  This implies an uncertainty in the
distance at time $t$,
\begin{equation}
\delta l(t) = t \delta v \gsim \left(\frac{\hbar}{m \delta l(0)}\right)
\left(\frac{l}{c}\right), 
\label{ineq2}
\end{equation}
where we have used $t/2 = l/c$ (and we have dropped an additive term
$\delta l(0)$ from the right hand side since its presence complicates the
algebra but does not change any of the results). Minimizing $(\delta
l(0) + \delta l(t))/2$, we get the quantum mechanical uncertainty relation
\begin{equation}
\delta l^2 \gsim \frac{\hbar l}{mc}.
\label{ineq3}
\end{equation}

Next, we make use of the principle of equivalence, by exploiting the 
equality of the inertial mass and the gravitational charge of the clock, to
eliminate the 
dependence on $m$ in the above inequality.  This will promote the quantum 
mechanical uncertainty relation to a quantum gravitational uncertainty 
relation, making the uncertainty
expression useful.  Let the clock at A be a
light-clock consisting of two parallel mirrors (each of mass $m/2$), a
distance of $d$ apart, between which bounces a beam of light.  On the one
hand, the clock must tick off time fast enough such that $d/c \lsim \delta
l/c$, in order that the distance uncertainty is not greater than $\delta
l$: $\delta l \gsim d$.
On the other hand, $d$ is necessarily larger than the Schwarzschild radius
$Gm/c^2$ of the mirrors ($G$ is Newton's constant) so that the time
registered by the clock can be read off at all:
$d \gsim \frac{Gm}{c^2}$.  From these two
requirements, it follows that
\begin{equation}
\delta l \gsim \frac{Gm}{c^2},
\label{ineq4}
\end{equation}
the product of which and Eq. (\ref{ineq3}) yields the (low-energy) quantum 
gravitational uncertainty relation\cite{karol} 
\begin{equation}
\delta l \gsim (l l_P^2)^{1/3},
\label{ineq5}
\end{equation}
where $l_P = (\frac{\hbar G}{c^3})^{1/2}$ is the Planck length. The
intrinsic uncertainty in space-time measurements just 
described can be interpreted as inducing
an intrinsic uncertainty in the space-time metric $g_{\mu \nu}$.
Noting that $\delta l^2 = l^2 \delta g$ and using Eq. (\ref{ineq5}) we get
\begin{equation}
\delta g_{\mu \nu} \gsim (l_P/l)^{2/3}.
\label{delg}
\end{equation}

The fact that there is an uncertainty in the space-time metric means that
space-time is foamy.  Eq. (\ref{ineq5}) and Eq. (\ref{delg}) constitute
our model of spacetime foam.  We note that even on the
size of the whole observable universe ($\sim 10^{10}$ light-years),
Eq. (\ref{ineq5}) yields a fluctuation
of only about $10^{-15}$ m.  We further note that, according to our
spacetime foam model,
space-time fluctuations lead to decoherence phenomena.  The point
is that the metric fluctuation $\delta g$ induces a multiplicative phase 
factor in the wave-function of a particle (of mass $m$)
\begin{equation}
\psi \rightarrow e^{i \delta \phi} \psi,
\label{wf}
\end{equation}
given by
\begin{equation}
\delta \phi = \frac{1}{\hbar} \int m c^2 \delta g^{00} dt.
\label{phase}
\end{equation}
One consequence of this additonal phase is that a point particle with mass 
$m > m_P$ ($m_P \equiv \hbar / c l_P$ is the Planck mass) is a
classical particle (i.e., it suffices to treat it classically).  

Though the fluctuations that space-time undergoes are extremely
small, recently
Amelino-Camelia has argued (convincingly, we think) that modern
gravitational-wave interferometers 
may soon be sensitive enough to test 
our model of space-time foam.\cite{amca} 
The idea is fairly simple. 
Due to the foaminess of space-time, in any distance
measurement that involves an amount of time $t$, there is a minute
uncertainty $\delta l \sim (ct l_{P}^2)^{1/3}$.  
But measuring minute changes in (the) relative distances (of the test
masses or the mirrors) is exactly what a gravitational-wave interferometer
is designed to
do.  Hence, the intrinsic uncertainty in a distance measurement for a time
$t$ manifests itself as a displacement noise (in addition to other sources
of noises) that infests the interferometers
\begin{equation}
\sigma \sim (ct l_{P}^2)^{1/3}.
\label{noise}
\end{equation}
We can write the displacement noise in terms of its Fourier transform, the
associated
displacement amplitude spectral density $S(f)$ of frequency $f$.  For a
frequency-band limited from below by the time of observation $t$, $\sigma$
is given in terms of $S(f)$ by\cite{radeka}
\begin{equation}
\sigma^2 = \int_{1/t}^{f_{max}}[S(f)]^2 df.
\label{spden}
\end{equation}
For the displacement noise given by Eq.
(\ref{noise}), the
associated $S(f)$ is 
\begin{equation}
S(f) \sim f^{-5/6} (c l_{P}^2)^{1/3}.
\label{SD}
\end{equation}
Since we are considering only the low-energy limit of quantum gravity,
we expect this formula for $S(f)$ to hold only for frequencies much smaller 
than the 
Planck frequency ($c/l_P$).  

We can now use the existing noise-level data\cite{abram} obtained at the
Caltech 40-meter interferometer to put a bound on $l_P$.  In particular, by 
comparing Eq. (\ref{SD}) with  
the observed noise level of $3\times
10^{-19} {\rm mHz}^{-1/2}$ near 450 Hz, which is the lowest noise level
reached
by the interferometer, we obtain the bound $l_{P} \lsim 10^{-29}$ m which,
of course,
is consistent with the known value $l_P \sim 10^{-35}$ m.
Since $S(f)$ goes like $f^{-5/6}$ according to Eq. (\ref{SD}), we can
look forward to the LISA generation of gravitational-wave
interferometers
for improvement by optimizing the performance at low frequencies.  (We hope
that the gain by going to lower frequencies will not be offset by other
factors such as a much
larger arm length of the interferometers.)

\section{Reply to the comments by Adler et al}

In this section we reply to the four points raised by Adler et al.
in their paper \cite{anos} "On the detectability of quantum spacetime foam
with gravitational-wave interferometers."

(1) Ref. \cite{anos} claims that if Wigner's clock is quantum mechanical
but not
free,
then the uncertainty limit becomes much smaller than
that (Eq. (\ref{ineq3})) obtained by Wigner and used by us.  In
particular, Adler et al. give the example of a quantum clock bound in a
harmonic oscillator potential.  These authors err in neglecting the fact
that the clock is then bound to something.  We can now consider that
something to be part of the clock (after all, we have already considered
the light emission and detection devices as part of our clock), and
proceed with our argument presented in Section II.

(2) Ref. \cite{anos} claims that Wigner's limit is based on another
unrealistic
assumption: that the clock does not interact with the environment.  In
particular, Adler et al. point out that, if the clock is sufficiently
large or complex, it will interact with its enviroment in such a way that
its wave function decoheres.  In addition, these authors claim, such
interactions may localize or "collapse" the wave function, resulting in
the clock wave function that does not spread linearly over macroscopic
times, as opposed to what we have used in Eq. (\ref{ineq2}).  We admit
that
Adler et al. have raised a good point.  But the question of wave function
decoherence in the context of fundamental spacetime measurements is quite
subtle.  We think it is much more reasonable that the phenomenon of
enviroment-induced decoherence is an outcome (rather than an input) of
quantum spacetime
measurements at the fundamental level, with gravity being the
universal agent of
quantum decoherence as argued in Section II and as emphasized by
us\cite{nvd1,nvd2}.

(3) Adler et al. observe that the existing noise-level data\cite{abram} 
obtained at the Caltech 40-meter interferometer can be used to place a
lower limit on the effective mass of the hypothetical clock in Eq.
(\ref{ineq3}).  They find the effective clock mass to be larger than 3
grams which, according to them, is such a remarkably large mass
that it
hardly seems plausible as a fundamental property of spacetime.  This time,
these authors err in forgetting that the length scale involved in the
Caltech 40-meter interferometer measurement is macroscopic and has nothing
to do with fundamental length scales.  To appreciate this point, one can
use Eq. (\ref{ineq3}) and Eq. (\ref{ineq5}) to show that the optimum mass
for Wigner's clock (optimum in the sense that it yields the smallest
uncertainty in distance measurements) is given by
\begin{equation}
m \sim m_P (l/l_P)^{1/3}.
\label{optm}
\end{equation}
Thus the optimum
mass of the quantum clock depends on $l$, the distance in
the distance measurement.  If $l$ is macroscopic, the optimum mass is much
larger than the Planck mass.  On the other hand, if we dare (recall that 
we expect our result to be valid only for the low-energy domain of quantum 
gravity) to use Eq. 
(\ref{optm}) in the measurement of a microscopic distance approaching the
fundamental length scale $l_P$, the optimum mass of the
hypothetical clock would approach $m_P$, the fundamental mass scale.
Therefore, the relatively large mass of the clock found in Ref.
\cite{anos} is to
be expected since the distance involved in the Caltech interferometer
measurement at 450 Hz is huge compared to the Planck length. 

The above three objections raised by Adler et al. are all directed at the
quantum uncertainty limit (Eq. (\ref{ineq3})) obtained by Wigner and used
by us.  There is a way, albeit an indirect one, to show that the
uncertainty limit (Eq. (\ref{ineq3})) actually should be quite palatable 
even to those who
believe that the intrinsic uncertainty in distance measurements is
independent of the distance being measured and is given simply by the
Planck length.  All it takes is to use Eq. (\ref{ineq3}) as the starting
point.  But for the bound on $m$, instead of Eq. (\ref{ineq4}), one uses 
\begin{equation}
l \gsim \frac{Gm}{c^2},
\label{cons}
\end{equation}
which is nothing but the mathematical statement of the obvious observation
that,
to measure the distance from A to B, point B should not be inside the
Schwarzschild radius of the clock at A.  Then one finds
\begin{equation}
\delta l \gsim l_P,
\label{Wheeler}
\end{equation}
the canonical uncertainty\cite{MTW} in distance measurements.  Thus the
only question remaining is whether the more restrictive bound on $m$ given
by Eq. (\ref{ineq4}) is also correct.  This brings us to the next comment.

(4) Adler et al. note that the presence of the measurement clock system
certainly produces a distortion of spacetime, but Eq. (\ref{ineq4}) tells
us that it also produces an uncertainty in spacetime distances of about
the same amount.  (This fact has not escaped our attention.  See Ref.
\cite{nvd1}
and \cite{nvd2}.)  They contend that Eq. (\ref{ineq4}) must be wrong.  In
particular, taking the spinning Earth as the quantum clock, they assert
that, with Eq. (\ref{ineq4}), one would conclude that objects in the 
vicinity of the Earth have a
minimum intrinsic position uncertainty of roughly the Schwarzschild radius
of the Earth, which is about 1 cm, and this is manifestly false by many
orders of magnitude.  Our reply is simply that one cannot use the spinning
Earth, excellent as it is as a clock for daily lives, as Wigner's quantum
clock in the
gedanken timing experiment.  For one thing, the spinning Earth, by itself,
cannot function as a clock.  One also needs the Sun or the stars, for
example, thus
complicating the already huge timing device.  As a clock, the spinning Earth
is not very accurate.  Let us imagine building a telescope with opening as
large as the earth.  For visible light, its resolving power is about
$10^{-14}$ radians.  The Earth and the telescope rotate by about $10^{-4}$
radians per sec.  Hence, the spinning Earth, as a clock, cannot be precise
beyond the $10^{-10}$ sec. level, which can be translated to yield a
distance measurement accuracy to about 1 cm, hardly the precision needed for
spacetime measurements at the fundamental level.  Intuitively, it is also
clear   
that a quantum spacetime measurement cannot
tolerate the use of a monstrously huge and massive clock like the spinning
Earth which causes such a distortion in the geometry of spacetime that it
completely overwhelms the uncertainty in distance measurements.  Recall
that even on the size of the observable universe, the end result of our
analysis yields a distance fluctuation of only about $10^{-15}$ m which is
much smaller than the Schwarzschild radius of the Earth.  It is true that
we have merely used the light-clock as a model clock and there may be
more ideal clocks to use; but due to its simplicity, the light-clock
fits the bill of a quantum clock for the gedanken timing experiment at the
fundamental level. 

\section{Comments on some other questions}

In this section, we comment on four more questions that we think some of
our readers may ask.

(1) In Section II, we require our light-clock to tick off time fast enough
such that $d/c \lsim \delta l/c$, implying that $d/c$ is the smallest unit
of
time for our light-clock.  Some readers may well ask whether it is not
possible
to have smaller units of time by taking fractions of $d/c$.  If it is
possible, then the inequality $d/c \lsim \delta l/c$ needs not hold.  Our
reply is that, to be accurate, $d/c$ is indeed the smallest unit of time
for our light-clock.  To make an analogy, one does not use a minute-clock
to time a
100-m dash which takes only about 10 sec, a fraction of a minute. 

(2) Recall that our use of the light-clock in the gedanken timing
experiment yields Eq. (\ref{ineq4}).  One may wonder if the ensuing
result (Eq. (\ref{ineq5})) is not just an artifact of our model
clock.  Thus it is logical to ask whether it is not possible to replace our
light-clock with some other types of clocks such that all those
inequalities (including the Schwarzschild bound) no longer hold.  In the 
absence of explicit examples, it is
hard to draw any conclusions.  But let us consider a clock made of a small
object revolving around a black-hole just outside its event-horizon.  (And
let us ignore the gravitational radiation problem.)
Timing is provided by the periodicity of the motion.  Then there is no
analog
of $d$, the separation of mirrors in the light-clock, and it follows that
those inequalities are no longer valid, so goes the hypothetical argument.
The trouble with this argument is that actually there is an analog of $d$,
given by the size of the orbit around the black-hole; and mass of the
clock here is that of the black-hole.  Therefore it follows that those 
inequalities in Eqs. (\ref{ineq4}) and (\ref{ineq5}) (as an order of
magnitude estimate) still hold.

(3) In Eq. (\ref{spden}), we have used $1/t$ as the lower limit of
integration; but what if the lower limit is actually a multiple (call it
n) of $1/t$?  The answer is that, since Eq. (\ref{noise}) holds only up to
a multiplicative factor of order 1, a short calculation shows that so long
as
the multiple n is no more than 2 orders off unity, Eq. (\ref{spden}) stands
as it is.

(4) One may worry that the metric fluctuations given by Eq. (\ref{delg})
yield an unacceptably large fluctuation in energy density.  Since we
asked ourselves this very question and answered it in Ref. \cite{nvd2} 
already, we
will be very brief here.  But let us generalize the discussion to metric
fluctuations of the form parametrized by $a$ with $0 < a \leq 1$
\begin{equation}
\delta g \gsim (l_P/l)^{a}
\label{gdelg}
\end{equation}
(corresponding to distance uncertainties of $\delta l \gsim l^{1-a}
l_P^{a}$ and displacement amplitude spectral densities of $S(f) \sim
c^{1-a} l_P^{a} f^{a-3/2}$). Models with larger spacetime
fluctuations are parametrized by
smaller values of $a$.
We note that, for our model of spacetime foam, $a=2/3$, while, for the
canonical model, $a=1$.  The case $a=1/2$ corresponds to the model 
of spacetime foam considered in Ref. \cite{GAC}. 
Regarding the metric fluctuation as a
gravitational
wave quantized in a spatial box of volume $V$, one finds\cite{nvd2} that
the energy density is given by
\begin{equation}
\rho \sim \left(\frac{m_P c^2} {V}\right),
\label{rho1}
\end{equation}
for $1/2 < a \leq 1$.  Thus the energy density associated with metric
fluctuations given by Eq. (\ref{gdelg}) is obviously small in
the large volume ($V >> l_P^3$) limit which we have assumed.  Note that
the energy density is of the form given by Eq. (\ref{rho1}) and holds, as
an order of magnitude estimate (consistent with what we have been using),
independent of the parameter $a$ so long as $a$ is not too close to 1/2.
For
$a=1/2$, one gets    
\begin{equation}
\rho \sim \left(\frac{m_P c^2}{V}\right) ln
\left(\frac{V^{1/3}}{l_P}\right).
\label{rho2}
\end{equation}
For $0 < a < 1/2$, one finds
\begin{equation}
\rho \sim \left(\frac{m_P c^2}{V}\right)
\left(\frac{V^{1/3}}{l_P}\right)^{1-2a}.
\label{rho3}
\end{equation}
The trend is clear: in general, larger spacetime fluctuations cost more 
energy.  Note that the energy density $\rho$ associated with metric
fluctuations (Eq. (\ref{gdelg})) is the smallest for the range of $a$ 
which includes the canonical model and our model of spacetime foam. 

\section{Conclusions}

In Ref. \cite{anos}, Adler et al. raise four objections to our work on
spacetime
measurements and spacetime foam.  Three of the objections are related to
the question whether
the quantum uncertainty limit obtained by Wigner and used by us is valid.
 These authors also criticize the gravitational uncertainty limit (Eq.
(\ref{ineq4})) obtained by us; they conclude that it is
an artifact of our choice of a particular type of hypothetical clock and
is, therefore, non-fundamental in nature.  While they have raised some good
points,
we believe their argument is flawed (as shown in Section III).  We agree 
that the question of an ideal quantum clock is not yet settled.  But it is
just inappropriate to use the spinning Earth as a quantum clock in a
fundamental spacetime measurement.  History has taught us that fundamental
physics is best explored with simple devices; our light-clock is a simple
device.

Since all the criticism by Adler et al. is related to issues of clocks,
perhaps a
better argument for a spacetime foam different from the canonical model is
one that does not use clocks.  As shown in Section IV, the energy density
associated with spacetime quantum fluctuations takes on the
smallest (and comparable)
values for those spacetime foam models with the parameter $a$ in the range
$1/2 < a \leq 1$ so
long as $a$ is not too close to 1/2.  So, it is possible that Nature chooses
to
have a larger spacetime fluctuation (than that predicted by the canonical
model) 
at a comparable cost of
energy.  This argument is very loose, but hopefully we have made our
point.  Only future experiments can tell which value of $a$ (i.e., which
spacetime foam model) Nature picks.  At present, if we assume that the 
distance uncertainty expressions given above are not off by more than
an order
of magnitude, a short calculation shows that the existing data provided by
the 
Caltech 40-meter
interferometer rule out models with $a < 0.54$.  We can expect more
stringent bounds on $a$ with modern gravitational-wave
interferometers.   

There is one theoretical consideration which sets our model of spacetime
foam ($a=2/3$) apart from the others.  It is its connection to the
holographic principle\cite{tHooft} which asserts that the number of degrees
of freedom of a region of space is bounded (not by the volume but) by the
area of the region in Planck units.  To see that, let us consider a region
of space with linear
dimension $l$.  According to the conventional wisdom, the region can be
partitioned into cubes as small as $l_{P}^3$.  It follows that the number
of degrees of freedom of the region is bounded by $(l/l_P)^3$, i.e., the
volume of the region in Planck units.  But according to our spacetime foam
model\cite{nvd1,nvd2,nvd3}, the smallest cubes inside that region have a
linear dimension of order $(l l_{P}^2)^{1/3}$.  Accordingly, the number of
degrees of freedom of the region is bounded by $[l/(ll_{P}^2)^{1/3}]^3$,
i.e., the area of the region in Planck units, as stipulated by the
holographic principle.
Thus one may even say that the holographic principle has its origin in the
quantum fluctuations of spacetime.  Evidence for our spacetime foam model
would lend experimental support for the holographic principle.

Finally we recall that spacetime (metric) fluctuations can be regarded as a
kind of quantized gravitational waves.  It is uncanny that, through future
refinements, modern
gravitational-wave interferometers like LIGO, VIRGO, and LISA, which are
designed to 
detect gravitational waves from neutron stars, supernovae, black-holes, and
the like, 
may also be able to detect, as a by-product, a very different 
kind of gravitational waves --- 
the kind that encodes the quantum fluctuations of spacetime.

\bigskip

\begin{center}
{\bf Acknowledgments}\\
\end{center}

One of us (YJN) thanks R. Weiss for a useful discussion.  This work was 
supported in
part by
the U.S. Department of Energy under \#DF-FC02-94ER40818 and
\#DE-FG05-85ER-40219,
and by the Bahnson Fund of the University of North Carolina at Chapel Hill.
Part of the work was carried out by YJN while he was on leave of absence at
MIT.  He thanks the faculty at the Center
for Theoretical Physics for their hospitality.

\end{document}